# Quantum key distribution based on time coding

T. Debuisschert, W. Boucher

*THALES Research and Technology, Domaine de Corbeville, 91404 Orsay Cedex, France*

**Abstract** : A quantum key distribution protocol based on time coding uses delayed one photon pulses with minimum time-frequency uncertainty product. Possible overlap between the pulses induces an ambiguous delay measurement and ensures a secure key exchange.

An alternative protocol to polarization coding[1] or phase coding[2,3] is proposed for quantum key distribution. It is based on a time coding technique which is expected to be robust against propagation medium disturbances. The information is coded on one photon pulses of duration T and with a chosen time delay with respect to a time reference. The possible delays are chosen by Alice so that possible pulses may overlap. Bob uses photon counters with a time resolution much better than the pulse duration and he measures the detection time with respect to the reference. He can perform only one measurement which may lead to an ambiguity on the delay evaluation.

The simplest configuration is a two states configuration in analogy with the B92 protocol[4]. Alice may send two kinds of pulses. One (e.g. bit 0) is coded with zero delay, the other one (e.g. bit 1) is coded with T/2 delay (Figure 1). The photon detection can occur within three different time intervals (Figure 1). The first one and the third one are non ambiguous and allow for an exact determination of the delay. The photon detection in the second time interval leads to an ambiguity on the delay determination. The spy Eve cannot deduce with certitude the delay chosen by Alice for all pulses she detects. She unavoidably introduces errors in the message when she sends back pulses of duration T to Bob.

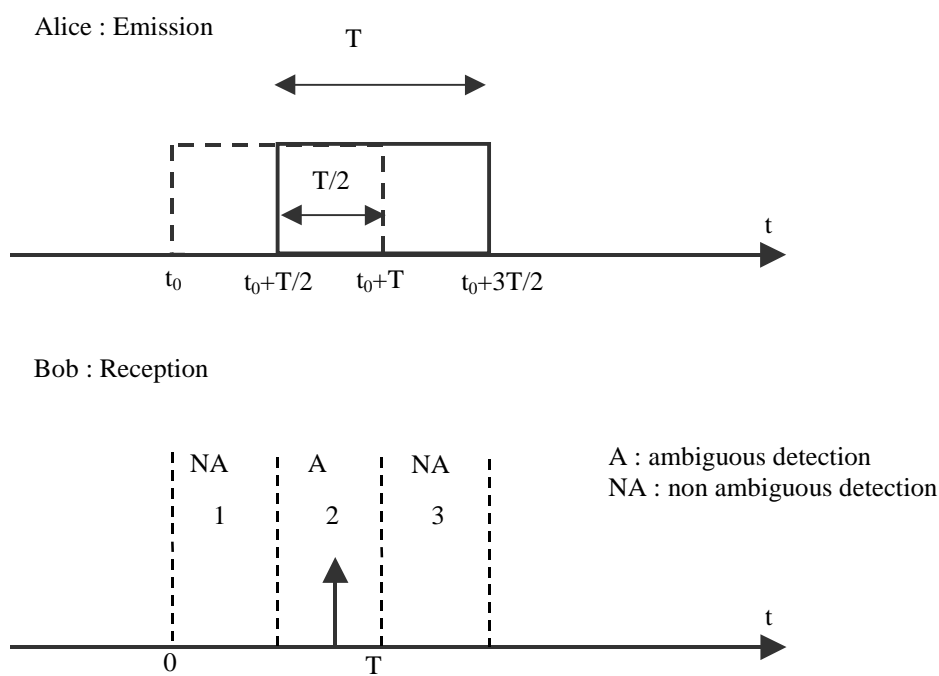

Figure 1 : Principle of the two states protocol. Alice sends pulses of duration T with chosen delay 0 or T/2. Bob measures the photon detection time. The interval 1 and 3 are non ambiguous and allow for delay determination. The interval 2 is ambiguous and does not allow for delay determination.

At this step, the transmission protocol is not yet quantum. Eve can get a perfect copy of the key after the reconciliation process. She only has to send back to Bob one photon pulses with a duration $T_E$ much smaller than T and with a delay identical to the one she measured. Bob cannot distinguish T pulses from $T_E$ pulses with only one measurement. To protect the transmission from that kind of attack, Alice sends pulses with minimum time-frequency uncertainty product, i.e. pulses with a coherence length equal to their duration. Bob sends at random the pulses he receives to a Mach-Zender interferometer with propagation time difference of T/2 and phase difference of $\pi$ between the two arms (Figure 2). The imbalance between the average photon number detected in each output arm of the interferometer varies with the pulse duration thus giving a way to measure that duration. The other arm of the input beamsplitter is sent to the photon counter which is used to establish the key between Alice and Bob (Figure 2).

This simple protocol can be extended to more than 2 states protocol. It can be implemented with standard optical components and photon counters. A first experimental demonstration is ongoing.

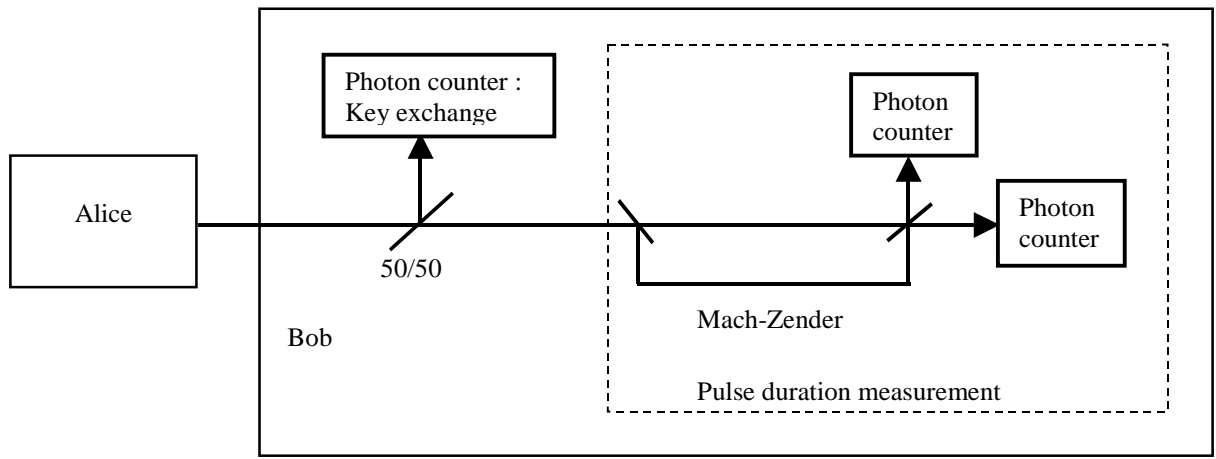

Figure 2 : Scheme of the experiment. The pulses sent by Alice are directed by Bob at random to a photon counter to establish the key, or to a Mach-Zender interferometer which allows for duration measurement of the pulses.

**References**

1. W.T. Buttler et al, Phys. Rev. Lett. 84, 5652 (2000)
2. P.D. Townsend et al, Electron. Lett. **29**, 634 (1993)
3. A.Muller et al, Appl. Phys. Lett. **70** (7), (1997)
4. C. H. Bennett, Phys. Rev. Lett. **68**, 3121 (1992)